\documentclass[aps,twocolumn,showpacs,preprintnumbers, floats]{revtex4}
\usepackage{amsfonts}
\usepackage{amsmath}
\usepackage{graphicx}
\usepackage{dcolumn}
\usepackage{bm}

\begin{document}

\preprint{}
\title{Local conductivity and the role of vacancies around twin walls of (001)-BiFeO$_3$ thin films}
\author{S. Farokhipoor and B. Noheda}
\affiliation{$^{1}$Zernike Institute for Advanced Materials, University of Groningen, Groningen 9747AG, The Netherlands}

\date{December 19, 2011}

\begin{abstract}
 BiFeO$_3$ thin films epitaxially grown on SrRuO$_3$-buffered (001)-oriented SrTiO$_3$ substrates show orthogonal bundles of twin domains, each of which contains parallel and periodic 71$^o$ domain walls. A smaller amount of 109$^o$ domain walls are also present at the boundaries between two adjacent bundles. All as-grown twin walls display enhanced conductivity with respect to the domains during local probe measurements, due to the selective lowering of the Schottky barrier between the film and the AFM tip (see S. Farokhipoor and B. Noheda, Phys. Rev. Lett. 107, 127601 (2011)). In this paper we further discuss these results and show why other conduction mechanisms are discarded. In addition we show the crucial role that oxygen vacancies play in determining the amount of conduction at the walls. This prompts us to propose that the oxygen vacancies migrating to the walls locally lower the Schottky barrier. This mechanism would then be less efficient in non-ferroelastic domain walls where one expects no strain gradients around the walls and thus (assuming that walls are not charged) no driving force for accumulation of defects.
\end{abstract}


\maketitle
\section{Introduction}

The importance of domain walls and their influence in the response of materials have been recognized since decades\cite{Kit1949} but it has been mainly magnetic domains walls, which typically extend to decades of unit cells, the ones most thoroughly investigated. Less widely known was, perhaps, that ferroelastic twin walls, spanning a few unit cells in width, are exceptional physical nano-objects with remarkable properties\cite{Sal1999,Sal2009}, being a notable example the observation of superconductivity at the ferroelastic domain walls of doped WO$_3$\cite{Sal1998}. Indeed, ferroelastic, or twin, walls are also inherent symmetry-breaking entities within a given material, with the potential to display the distinct additional physical responses associated to that fact. Moreover, they also represent areas of intense strain gradients that can attract large concentrations of dopants or defects, which will also modify greatly and very locally the materials' properties\cite{Sal2005}. In addition, the strain gradients associated to those walls can induce electrical dipoles, due to the so-called flexoelectric effect\cite{Kog1964,Cro2006,Ma2008,Cat2011}.

Even if tantalizing results on twin walls were known and twin walls had been accessed and imaged \cite{Sal2004,Shi2004}, it is only recently that it is made clear that, using thin film deposition techniques, reproducible control and tunability of the type, orientation and periodicity of domain walls, separated by distances of only tens to hundreds of nanometers, is possible. Thus, also due to the widespread use of local probes, it is now realistic to think of using the domain walls and their unique properties as devices\cite{Sal2010,Bea2009}.

But what has boosted the general interest on domain walls is the recent discovery that domain walls of multiferroic (ferroelectric and antiferromagnet) BiFeO$_3$ thin films are considerably more conducting than the domains and, thus, provide well-defined, local paths of conduction through the thin films\cite{Sei2009}. BiFeO$_3$ thin films are monoclinic\cite{Shi2005} but, in fact, they are quasi-rhombohedrally distorted and, thus, the ferroelastic domain walls that they present are very similar to those expected in a rhombohedral perovskite, that is 71$^o$ and 109$^o$ domain walls\cite{Str1998}. Initially, conducting behavior was observed only in artificially-written 109$^o$ domain walls\cite{Sei2009}. However, it was later reported that also as-grown 71$^o$ walls displayed enhanced conductivity with respect to the domains\cite{Chi2010}, although that was one order of magnitude smaller than the conductivity found in as-grown 109$^o$ domain walls. This seems to be in agreement with the theoretical calculations that assign the origin of conduction to a decrease in the band gap of the material at the domain walls\cite{Chi2010,Sei2009}. However, in samples grown under different conditions, we have observed that conductivity at 71$^o$ domain walls can be as large as that in 109$^o$ domain walls and that its main origin is the lowering of the Schottky barrier between the BiFeO$_3$ n-type semiconducting film and the metallic top electrode\cite{Far2011}. The predicted reduction of the band gap at the walls would then appear as a secondary effect, responsible for subtler changes in conduction.

Therefore, if we want to achieve control of walls either as devices or as parts of a device, it is important to investigate the mechanism or mechanisms for selective conduction through domain walls and, in case that there is more that one possibility, to find out which one is most robust and convenient. This becomes now essential because, according to recent reports, conduction at domain walls is a rather general phenomenon\cite{Guy2011}, only the most relevant mechanisms seem to vary. As one can expect, different conduction mechanisms are likely to have different relevance in different types of walls. Fowler-Nordheim (FN) tunneling has been put forward as the mechanism for conduction in artificially-written 109$^o$ domain walls\cite{Sei2010,Mak2011}. Recent work has also focused on walls between ferroelectric 180$^o$ domains, which do not involve strain gradients\cite{Guy2011,Mak2011b} and indeed a different conduction mechanism (bulk-limited Poole-Frenkel emission) has been observed in this case. Similarly, charged walls show again different conduction characteristics\cite{Eli2011}. Later work has gone further in exploring the conduction involved in truly one-dimensional conduction paths defined by vortex states in ferroelectrics, around which both charged walls and large strain gradients coexist\cite{Bal2011}

In this paper we focus on ferroelastic domain walls of BiFeO$_3$ thin films to complement the results reported in ref.\cite{Far2011}. We include here a more detailed description of our samples and a more detailed analysis of the different conduction mechanisms that have been considered and the reasons why they have been discarded, as well as how the Schottky barriers have been calculated. We have also pointed out\cite{Far2011} that the lowering of the barrier at the walls is likely to be induced by migration of charged defects to the walls (as it is common in other perovskites\cite{Sal2005}), since that would create a potential step at the interface\cite{Sze}. We postulated that the charged defects are oxygen vacancies, based on the observed enhancement of current in samples with increased oxygen vacancy concentration. In this paper we show evidence of the crucial role played by the oxygen content in determining the domain wall conductivity.

\section{Experimental}

BiFeO$_3$ thin films were grown by pulsed laser deposition (PLD) assisted by RHEED using a system designed by Twente Solid State Technology (TSST)\cite{TSST} provided with an excimer laser (Lambda Physik COMPex Pro 205) filled with KrF to produce a wavelength of 248 nm. The operating laser frequency and fluence were 0.5Hz and 2 J/cm$^2$, respectively. The chamber was evacuated up to a background pressure of 10$^{-8}$ mbar. The films were grown at a temperature of 670$^oC$ in an oxygen pressure of 10$^{-1}$ mbar. After growth, the oxygen pressure was increased to 100mbar and the films were cooled down at a rate of 3$^o$ C/min, except in particular cases in which we intended to investigate the effect of different oxygen vacancy content, for which the cooling rate was decreased(increased) or the oxygen pressure was increased(decreased). The films are grown on single-terminated (TiO$_2$)(001)-oriented SrTiO$_3$ substrates\cite{Kos1998} covered by a buffer electrode layer of SrRuO$_3$. The 5nm thick SrRuO$_3$ layer was deposited also by PLD immediately before the BiFeO$_3$ deposition, with a substrate temperature of 600$^o$C, an oxygen pressure of 0.13 mbar and laser frequency and fluence of 1 Hz and
2J/cm$^2$, respectively.

\begin{figure}[h]
    \centering
         \includegraphics[width=7cm]{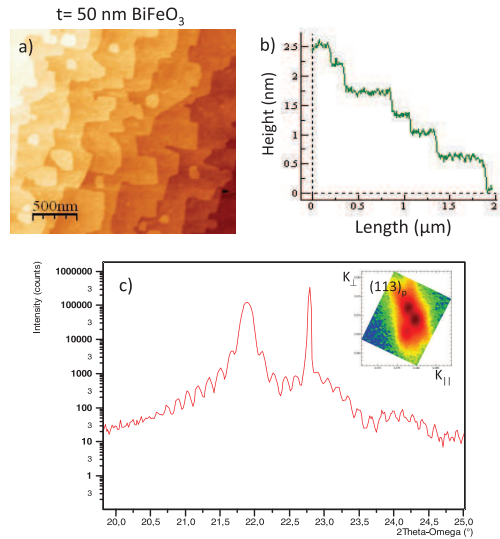}
   \caption{a) 2$\mu$m x 2$\mu$m atomic force microscope (AFM) image of the topology of a 50nm thick BiFeO$_3$ film grown by PLD on a 5nm thick SrRuO$_3$ electrode on a low miscut, single-terminated, (001)-oriented, SrTiO$_3$ substrate. b) A linear scan of such image shows atomically flat terraces following the substrate steps. c) XRD pattern of the same film around the (001) reflection of the BiFeO$_3$ film (low-angle peak) and SrTiO$_3$ substrate (high-angle peak) The thickness fringes corresponding to both the film and the electrode are clearly visible and indicate the excellent quality of the interfaces. In the inset, a reciprocal space map around the pseudo-cubic (113) reflection of the film shows that it is monoclinic with the [001] direction parallel to that of the substrate\cite{Dau2009}.}
   \label{fig:Fig1}
\end{figure}

(001)-oriented BiFeO$_3$ films with thickness ranging between 40 and 68 nm are used in this study. These films show atomically flat terraces (see Figure 1a-b, flat interfaces and good crystal quality (Figure 1c). They form monoclinic, pseudo-rhombohedral, domains with the polarization very close to the pseudo-cubic [111]$_p$ direction. From the eight possible polarization orientations, only the four down oriented domains are present in these films, in agreement with other reports\cite{Jan2009,Ho2008}, and thus no 180$^o$ domain walls are present. The observed domains are organized in bundles\cite{Ivr2010} of parallel 71$^o$ domain walls\cite{Dau2009}. Adjacent bundles are oriented orthogonal to each other and at the boundaries in between bundles some 109$^o$ domain walls are also found.

From the reciprocal space maps around non-specular reflections (inset of Figure 1c and ref.\cite{Dau2009}), it can be inferred that the (001) atomic planes of the film are parallel to those of the substrate and buffer layer
and, thus, that there is no out-of-plane tilt or buckled interface, as reported for some (001)-oriented rhombohedral
perovskites\cite{Jan2009,Liu2010,Mar2001}. This is important because the nature of the interfaces will largely influence the conduction properties of the films\cite{Pin2009}

 We have performed DC transport measurements on a VEECO (now Bruker) \emph{Dimension V} Conductive Atomic Force
Microscope (CAFM) provided with a (Co-Cr coated Si) metallic tip. Electrically, the tip is connected to the ground of the microscope and a bias voltage is applied to the metallic sample holder, which is connected to the SrRuO$_3$ bottom electrode using silver paste on the side of the sample. A so-called TUNA$^{TM}$ amplifier allows to measure currents through the film in the range of 100 fA-10$\mu$A, using four different gain factors (and sensitivities). Measurements are performed both as a function of temperature and applied voltage in the range of -5 V to +5V, well below the voltages at which currents associated with domain switching are observed (at about 7 V). The temperature was increased in steps of 10 K in ambient conditions from room temperature up to 155 $^o$C. In order to prevent spurious transient currents that are related to ionic conductivity, we included a delay time of several minutes from the application of the voltage to the starting of the measurements.

\section{Results and discussion}

Figure 2c shows localized areas of enhanced conduction around both 71$^o$ and 109$^o$ domain walls in the as-grown
(001)-BiFeO$_3$ films described above, as we have previously reported\cite{Far2011}. Similar results have been observed by Chiu et al.\cite{Chi2010} using scanning tunneling microscopy. However, in our films, unlike in those of ref.\cite{Chi2010}, the magnitude of the current in the 71$^o$ walls is as large as that of the 109$^o$ walls. This difference between the two studies could be due to the nature of the interfaces (out-of-plane twinning versus in-plane twinning)\cite{Liu2010,Pin2009}, which may invoke different conduction mechanisms, as mentioned in the previous section. Moreover, the 71$^o$ domain walls in our investigated films showed a somewhat more (statistically) consistent behavior than that of the 109$^o$ walls, probably due to the fact that the later are located at the edges of the orthogonal bundled regions, where the strain is
enhanced, possibly inducing dislocations or other defects. Therefore, we have focussed most of our studies on the 71$^o$ domain walls.

\begin{figure}
        \includegraphics[width=7cm]{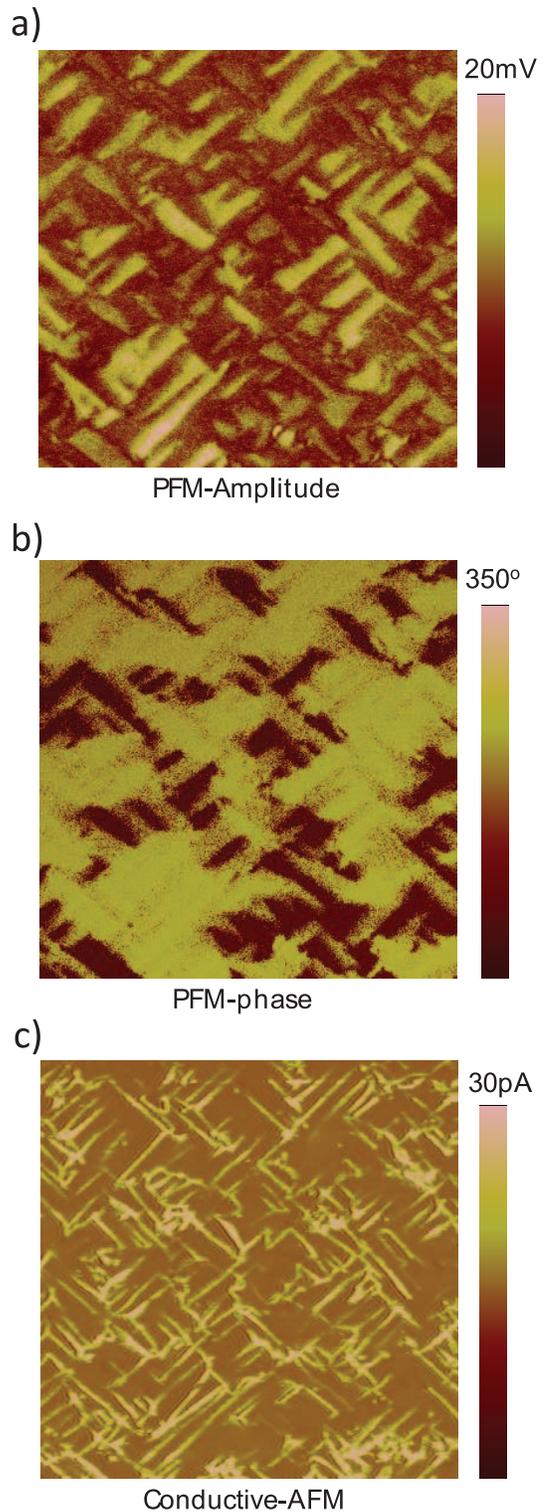}
        \caption{Piezo-force microscopy (PFM) amplitude (a) and phase (b) images of a BiFeO$_3$ film as those described in the text. Both images are needed to determine the type of domain walls. c) Conductive-AFM (C-AFM) image of the same area of the film showing enhanced currents at the domain walls. Both 71$^o$ and 109$^o$ domain walls show similar magnitudes of the current.}
    \end{figure}

The magnitude of the measured current at the walls is of about 10 pA at 4 V at room temperature and can increase up to 3nA
for temperatures around 150$^o$C. By measuring I-V curves as a function of temperature (above room temperature), we have
evaluated the three most likely mechanisms leading to conduction through a metal-semiconductor-metal system, as well as the
possibility of space charge limiting the current. We have considered a linear dependence between the applied voltage and the electric field, which is not obvious giving that one of the electrodes is a sharp tip. However, recently, Guyonnet et al.\cite{Guy2011} have shown that this is, indeed, a good approximation.

\begin{figure*}
         \includegraphics[width=15cm]{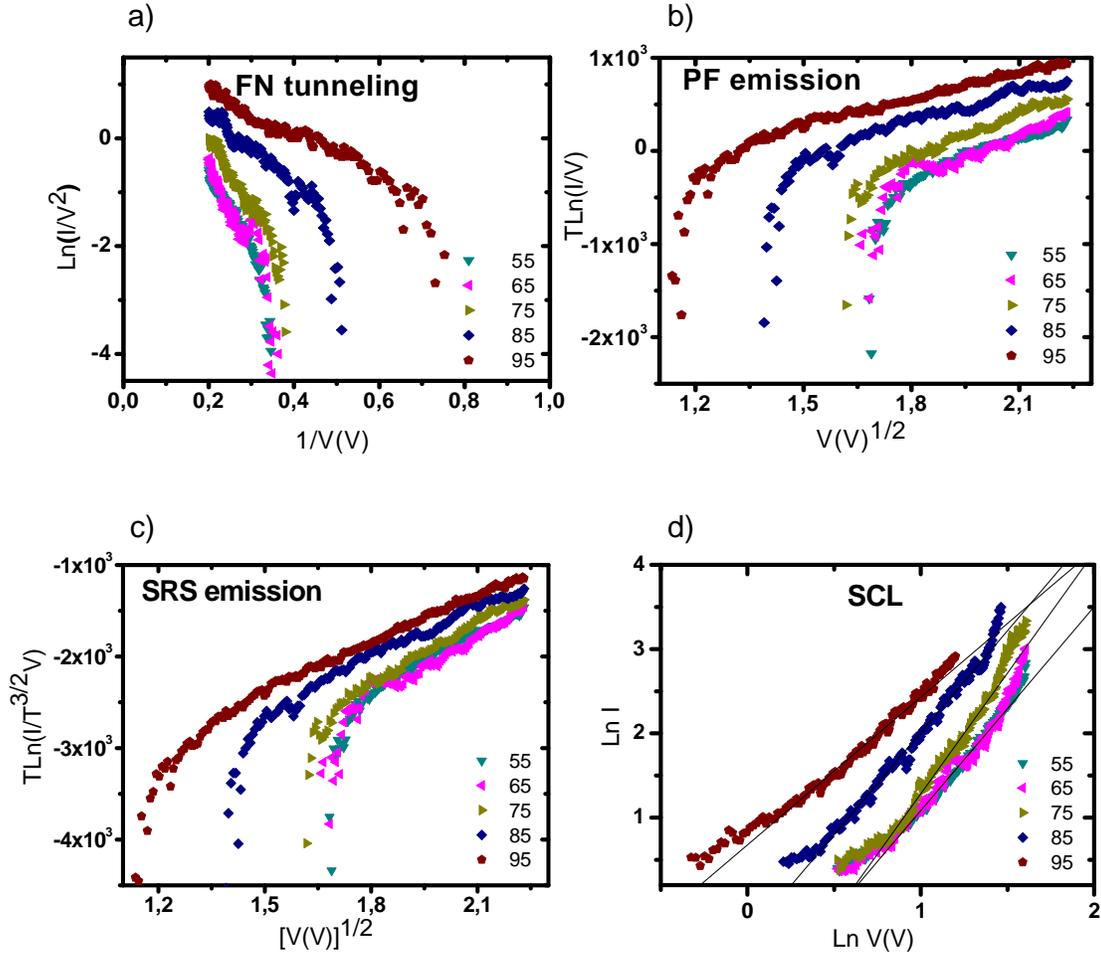}
  \caption{The measured I-V curves for different temperatures at 71$^o$ domain walls are plotted with different axes in order to make the curves linear and temperature-independent for a) Fowler-Nordheim (FN) tunneling; b) Poole-Frenkel (PF) emission; c) Simmons-Richardson-Schottky (SRS) emission and d) space charge limited (SCL) current. a) shows temperature dependent slopes that are inconsistent with the FN mechanism; the slope in b) leads to a too small optical dielectric constant of 0.4$\pm$0.1; while the slope of c) gives the right value of 6.5$\pm$1.5. d) shows the data for voltages below 3.7 V, which deviate from linearity in figure 3c). Power laws with exponents close to 2 seem to fit the data reasonable well\cite{Far2011} in agreement with a space charge limited regime.}
 \end{figure*}

Figure 3a) represents the data in coordinates related to Fowler-Nordheim tunneling, that is ln(I/V$^2$) is plotted against
1/V for different temperatures. Fowler-Nordheim tunneling should be temperature-independent and, in these coordinates, should show a linear dependence. Figure 3a shows quite linear behavior mainly in the 3-5V region but there is a clear temperature dependence that indicates that this is not the relevant mechanism for conduction.

Another possible mechanism is Poole-Frenkel (PF) emission, by which current is supported by thermal emission from trapped carriers in the bulk of the semiconductor, where the trap barrier is lowered by the electric field. This has been reported to be the main conduction mechanism in some ferroelectric oxides, including BiFeO$_3$ \cite{Pin2009} and
PZT\cite{Guy2011}. The field and temperature dependence of the PF currents are similar to those of the Richardson-Schottky (RS) emission, a third mechanism to be considered. The best way to distinguish between the two of them is by extracting the
high-frequency dielectric constant from the slopes of the linearized plots and see in which case it agrees with that known for the material.\cite{Sim1967}. In addition, the Richardson-Schottky equation is only strictly correct in the case of semiconductors with relatively small electronic mean free paths; while for insulators the modification proposed by Simmons (which we will call Simmons-Richardson-Schottky (SRS) emission)\cite{Sim1965} should apply.

\begin{figure}
         \includegraphics[width=7cm]{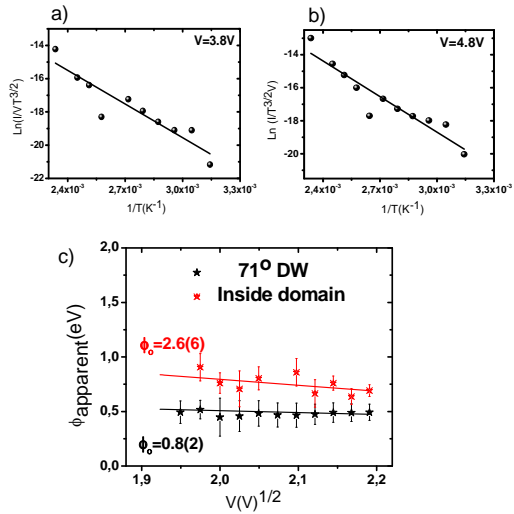}
        \caption{Ln(I/(VT$^{3/2}$) \emph{versus} T$^{-1}$ for two particular voltages, V= 3.8 V (a) and V= 4.8V (b), limits within which the SRS mechanism fulfills. The slopes of those curves give rise to what has been called $\phi$$_{apparent}$\cite{Pin2009}. c) According to the SRS mechanism, $\phi$$_{apparent}$ vs. V$^{1/2}$ should show linear behavior with the intercept at the origin being the height of the Schottky barrier ($\phi$$_{o}$). The $\phi$$_{o}$ values extracted in this way for domains and domain walls are indicated in the plot.}
\end{figure}

Thus, in Fig 3 b)-c), we plot the data in the coordinates that make the current data due to PF emission or SRS emission linear and temperature-independent (Tln(I/V) \emph{versus} V$^{1/2}$ and Tln[I/(VT$^{3/2}$)] \emph{versus} V$^{1/2}$, respectively). In this way, in both plots, all the curves (independent of temperature) should show the same intercept and the same slope, which is inversely-related to the dielectric constant\cite{Sim1967}. The linearity of the fits is considerably improved compared to the FN tunneling fit, in particular at higher temperatures and higher voltages. One can see that the slopes are in both cases (PF as well as SRS) temperature-independent in a very good approximation. The
collapse of the intercept seems better in the SRS case (leading to a well-defined Schottky barrier). However, it is the value of the optical dielectric susceptibility extracted from the slopes what makes us conclude that Schottky emission (SRS) is the pertinent conduction mechanism in this case: In the case of PF emission a value of 0.4$\pm$0.1 for the optical dielectric constant is found from the slopes of Fig 3b) for 71$^o$ domain walls and a value of 0.16$\pm$0.06 is found for the domains (figure not shown). For SRS emission, a dielectric permittivity of 6.5$\pm$1.5 is found at the 71$^o$ domain walls (from figure 3c), in perfect agreement with the expected value of 6.25 (see e.g.\cite{Pin2005}). A value of 3.6$\pm$0.1 is found at 109$^o$ domain walls; while a worse agreement of 1.8$\pm$0.3 is obtained for the domains. It is worth to mention that, due to the low values of the current inside the domains, these data are very noisy and only a few higher temperature curves could be analyzed.

To calculate the Schottky barrier height we follow ref. \cite{Pin2007}: we first plot $Ln(I/{VT^{3/2}})$ versus $1/T$. A linear fit should be obtained for the proper voltage regime (voltage large enough to support the Schottky emission mechanism, in our case above 3.8V). Figures 4 a)-b) show the fits for two specific (the largest and the lowest) voltages. $\Phi_{apparent}$ is defined as the slope of these linear fits. This is the exponential term in the Simmons-Richardson-Schottky emission\cite{Sim1965}: (-$\Phi_{o}$/k$_B$)+(e/k$_B$)(eV/d$\pi$$\varepsilon_{o}$ K)$^{1/2}$, where $\Phi_{o}$ is the Schottky barrier, k$_B$ is the Boltzmann constant, $\varepsilon_{o}$ is the permittivity of free space, K is the dielectric constant, V is the applied DC bias and d is the thickness of the film. Next, we plot the slopes \emph{versus} V$^{1/2}$ (Figure 4c). The intercept of the obtained graph (at V=0) is the Schottky barrier height. The barrier height for 71$^o$ domain walls is determined to be (0.8 $\pm$ 0.1) eV, however, inside domains this value changes to (2.6$\pm$0.7) eV.

\begin{figure}
         \includegraphics[width=7cm]{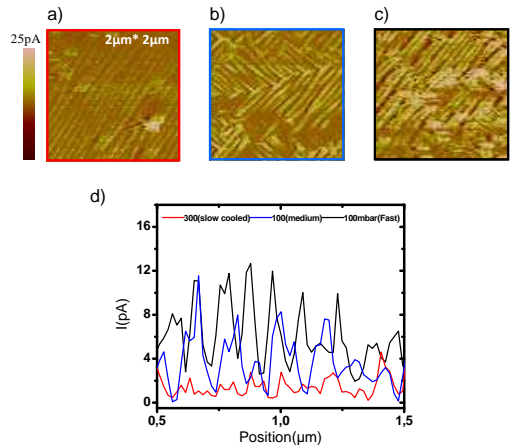}
        \caption{C-AFM images on an 2$ \mu$mx 2$ \mu$m area for three samples grown under identical conditions but cooled down after growth a) under an oxygen pressure of PO$_2$= 300 mbar and a cooling rate of 2K/min; b) under PO$_2$= 100mbar and cooling rate of 3 K/min and c)under PO$_2$= 100mbar and cooling rate of 30-4 K/min. The current scale is the same for the three images d) Linear scans across 1 $\mu$m of the previous images are shown in red (for a)), blue (for b) and black (for c).}
\end{figure}

 One can see in Fig. 3b-c) that at the lower voltages the data deviates from the linear trend indicating a different behavior. The lower the temperature, the higher the voltage at which the SRS regime crosses over to this low voltage regime. It is known that the general signature of space charge limited (SCL) conduction in an insulator of thickness $d$ (with parallel-plate electrodes) is I$\propto$ V$^n$/$d$$^{n+1}$, where V is the applied voltage and
$n$$\geq $2 (n= 2 in the absence of traps)\cite{Ros1955}. Typically, the SCL regime is found at the higher voltages (once there are enough carriers in the insulator to build up space charge). However, charged defects or, in a ferroelectric material, the polarization surface charges themselves can behave similarly to the space-charge regions and could lead to a similar I-V dependence\cite{Blo1994}. In Fig. 3d) we plot the low voltage data that, according to Figure 3c), deviates from the linear Schottky emission behavior. As discussed before, this deviation is more frequently observed at the lower temperatures and it is in reasonable agreement with an n= 2 slope. In this low voltage regime (2.5-3.5V), Arrhenius behavior of the current has been observed, with a thermal activation energy of 0.73(9)eV\cite{Far2011}. This value is consistent with trapped electrons from clusters of oxygen vacancies in perovskites.\cite{Sei2010}.

That oxygen vacancies play a very important role in controlling the conduction in these films has been clearly observed by making samples with different oxygen contents, either by changing the cooling rate or the oxygen pressure after growth, but otherwise growing them under identical conditions. These results are summarized in Figure 5, where three C-AFM images are displayed for samples with low, medium and high vacancy content. An increasing amount of current is obtained with decreasing oxygen content. A clear dependence of the current with growth oxygen pressure was also observed by Seidel et al.\cite{Sei2010} in 109$^o$ twin walls.

All this brings forward the following scenario: the conduction observed in the low voltage regime is coming from thermally excited electrons from defect states (possibly related to oxygen vacancies) located close to the bottom
of the conduction band. With increasing voltages, oxygen vacancies move towards the surface and lower the Schottky barrier with the top electrode, which eventually allows large conduction through the films. This effect is, however, not homogeneous across the film: strain gradients associated to ferroelastic domain walls create an inhomogeneous distribution of oxygen vacancies and the selective reduction of the Schottky barrier around the walls. The mechanism for conduction is thus expected to be different in non-ferroelastic domain walls.

To conclude, we have observed that in (001)- BiFeO$_3$ thin films grown on SrRuO$_3$-buffered (001)-SrTiO$_3$ substrates, the currents above about 3.8 V (for films with thickness of about 50nm) are determined by the Schottky barrier between the BiFeO$_3$ film and the Co-Cr tip. This barrier is largely decreased in the domain walls with respect to the domain, causing the observed conduction enhancement at the walls. Oxygen vacancies are observed to play a crucial role in this. At low voltages the current behaves like it is limited by space charge. The importance of the interplay between the surface polarization charges and the charged defects have been recently put forward in BiFeO$_3$ single crystals by\cite{Yi2011} and in films\cite{Bal2011,Eli2011}. This interplay is, of course, very different at the domain and at the walls and it has been investigated in relation with domain/domain wall conductivity in the present BiFeO$_3$ thin films. The results will be discussed elsewhere\cite{Far2012}.

Useful discussions with Tamalika Banerjee, Sergei Kalinin, Patrycja Paruch, R. Ramesh, James F. Scott and Pavlo Zubko are gratefully acknowledged. This work is part of the research programme on Functional Nanowalls (TOP Grant), which is financed by the Netherlands Organization
for Scientific Research (NWO).

\end{document}